\documentclass[]{spie}  

\usepackage{amsmath,amsfonts,amssymb,upgreek}
\usepackage{graphicx}
\usepackage[colorlinks=true, allcolors=blue]{hyperref}

\title{A High-Resolution Spectrograph for the 72cm Waltz Telescope at Landessternwarte, Heidelberg}

\author[a]{M. Tala}
\author[a]{P. Heeren}
\author[a]{M. Grill}
\author[a]{R.J. Harris}
\author[b]{J. St\"urmer}
\author[c]{C. Schwab}
\author[d]{T. Gutcke}
\author[a]{S. Reffert}
\author[a]{A. Quirrenbach}
\author[a]{W. Seifert}
\author[a]{H. Mandel}
\author[a]{L. Geuer}
\author[a]{L. Sch\"affner}
\author[e]{G. Thimm}
\author[f]{U. Seemann}
\author[a]{J. Tietz}
\author[a]{K. Wagner}
\affil[a]{Landessternwarte, Zentrum f\"ur Astronomie der Universit\"at Heidelberg, K\"onigstuhl 12, 69117 Heidelberg, Germany}
\affil[b]{Department of Astronomy and Astrophysics at University of Chicago, 5640 S Ellis Avenue, Chicago IL 60637, USA}
\affil[c]{Department of Physics and Astronomy at Macquarie University, Sydney, NSW 2109, Australia}
\affil[d]{Max-Planck-Institut f\"ur Astronomie, K\"onigstuhl 17, 69117 Heidelberg, Germany}
\affil[e]{Astronomisches Rechen-Institut, Zentrum f\"ur Astronomie der Universit\"at Heidelberg, M\"onchhofstra$\ss$e 12-14, 69120 Heidelberg, Germany}
\affil[f]{Institut f\"ur Astrophysik, Friedrich-Hund-Platz 1, 37077 G\"ottingen, Germany}

\authorinfo{Further author information: (Send correspondence to M. Tala)\\M. Tala: E-mail: m.tala@lsw.uni-heidelberg.de, Telephone: +49 152 065 58325\\ }

\begin{document} 
\maketitle

\begin{abstract}
     The Waltz Spectrograph is a fiber-fed high-resolution \'echelle spectrograph for the 72 cm Waltz Telescope at the Landessternwarte, Heidelberg. It uses a 31.6 lines/mm 63.5$^{\circ}$ blaze angle \'echelle grating in white-pupil configuration, providing a spectral resolving power of $R\sim$65,000 covering the spectral range between 450$-$800\,nm in one CCD exposure. A prism is used for cross-dispersion of \'echelle orders. The spectrum is focused by a commercial apochromat onto a 2k$\times$2k CCD detector with 13.5$\upmu$m per pixel. An exposure meter will be used to obtain precise photon-weighted midpoints of observations, which will be used in the computation of the barycentric corrections of measured radial velocities. A stabilized, newly designed iodine cell is employed for measuring radial velocities with high precision. Our goal is to reach a radial velocity precision of better than 5 m/s, providing an instrument with sufficient precision and sensitivity for the discovery of giant exoplanets. Here we describe the design of the Waltz spectrograph and early on-sky results. 
\end{abstract}


\section{INTRODUCTION}
\label{sec:intro}  
We have been carrying out a Doppler survey for exoplanets around giant stars since 1999 at Lick Observatory, using the Hamilton Echelle Spectrograph{\cite{Vogt87}} with an iodine cell{\cite{Butler1996}} coupled to the 60cm Coude Auxiliary Telescope (CAT). Planets in six systems have been published so far; several more will follow in the near
future {\cite{Frink2001}$^,$\cite{Reffert2006}$^,$\cite{Quirrenbach2011}$^,$\cite{Mitchell2013}}. Moreover, we have analyzed the giant planet occurrence rate in our sample as a function of mass and metallicity{\cite{Reffert2015}}. We found that the planet occurrence rate correlates with metallicity as for main-sequence stars, and that there is a maximum in the giant planet occurrence rate for stellar masses around 1.9 M$_{\odot}$, whereas the giant planet occurrence rate quickly decreases for more massive stellar hosts.\\
The Lick Doppler survey of giant stars ended when the iodine cell at Lick was damaged in 2011 {\cite{Fischer2014}}. Since then, we have been looking for alternatives to continue our survey, especially to confirm long-term trends that we see in our data which hint at sub-stellar companions with periods of a decade or more. The 72cm Waltz Telescope of the Landessternwarte (LSW), located on top of the K\"onigstuhl Observatory in Heidelberg, Germany, seems perfectly suited for the task if equipped with a high resolution spectrograph. It compares well with Lick in several respects: the telescope and the throughput of the spectrograph are slightly larger than at Lick, leading to shorter exposure times. The weather and sky conditions are worse, but if we use it every good night we estimate that we can get as many or slightly more spectra than we got at Lick to continue our survey with the same cadence as before, 5 nights/month. At the same time as carrying out the scientific survey, the telescope and spectrograph will also be used for student education and thus serves a double purpose.\\
Here we describe the design of a new precision radial velocities (RV) instrument, the Waltz Spectrograph. The instrument covers the spectral range between 450$-$800\,nm and provides a resolving power of $\sim$65,000. A first prototype was constructed based on a layout with a grating in off-plane angle; this instrument was a replica of the Yale Doppler Diagnostic Facility (YDDF) laboratory spectrograph at Yale{\cite{Schwab2012}}. We then modified the design to a full white-pupil layout for improved efficiency. It will be fed by a rectangular optical fiber, providing mechanical and illumination stability. The spectrograph is installed in a separate room, to minimize temperature fluctuations that lead to mechanical flexures. An image slicer will be placed on the front end to transform the image of the star into a pseudo slit to couple star light into the optical fiber with high efficiency. We use an iodine cell in the telescope beam as a stable reference for the measurement of high precision radial velocities. A Th-Ar hollow cathode lamp can be used as the default wavelength calibration source when not working in the high precision RV regime, by observing a Th-Ar spectrum before and after each observation. Here we describe the optical and mechanical design of the instrument, as well as early results obtained in the optics laboratory of LSW.
\begin{table}[ht]
\caption{The main features of the Waltz Spectrograph.} 
\label{tab:parameters}
\begin{center}       
\begin{tabular}{|l|l|} 
\hline
\rule[-1ex]{0pt}{3.5ex}  R ($\lambda/\Delta\lambda$)  & $\sim$65 000  \\

\rule[-1ex]{0pt}{3.5ex}  Wavelength coverage & 450 $-$ 800 nm   \\

\rule[-1ex]{0pt}{3.5ex}  Aperture on sky & 2.5 arcsec  \\

\rule[-1ex]{0pt}{3.5ex}  Fiber-feed & Rectangular fiber, 25\,$\times$100$\mu$m size  \\

\rule[-1ex]{0pt}{3.5ex}  Collimator & Parabolic mirror, F\,=\,900\,mm, D\,=\,300\,mm  \\

\rule[-1ex]{0pt}{3.5ex}  Echelle grating & 31.6 grooves/mm, 63.5$^{\circ}$ blaze angle  \\

\rule[-1ex]{0pt}{3.5ex}  Cross-dispersion prism & Schott F2 glass, 60$^{\circ}$ apex  \\

\rule[-1ex]{0pt}{3.5ex}  Camera &Apochromatic lens, F = 530\,mm, D = 106\,mm  \\
\rule[-1ex]{0pt}{3.5ex}  Detector & Andor iKon-L 936, 2k$\times$2k, 13.5$\upmu$m per pixel \\

\rule[-1ex]{0pt}{3.5ex}  Average spectral sampling & 2.6 pix  \\

\rule[-1ex]{0pt}{3.5ex}  Instrument efficiency (only spectrograph) & $\sim$46.5\% at 633\,nm   \\



\hline
\end{tabular}
\end{center}
\end{table}

\section{Instrument description}
The Waltz Spectrograph will be installed at the 72 cm Waltz Telescope located at Landessternwarte in Heidelberg, Germany. This telescope is a F/18.6 reflector with a plate scale of 15.03 arcsec/mm used in Nasmyth focus. The instrument consists of three different modules: the front end module, the calibration unit and the spectrograph. These are connected with optical fibers. In the following sections we will describe each one of these modules.
 \begin{figure} [ht]
   \begin{center}
   \begin{tabular}{c} 
   \includegraphics[height=8cm]{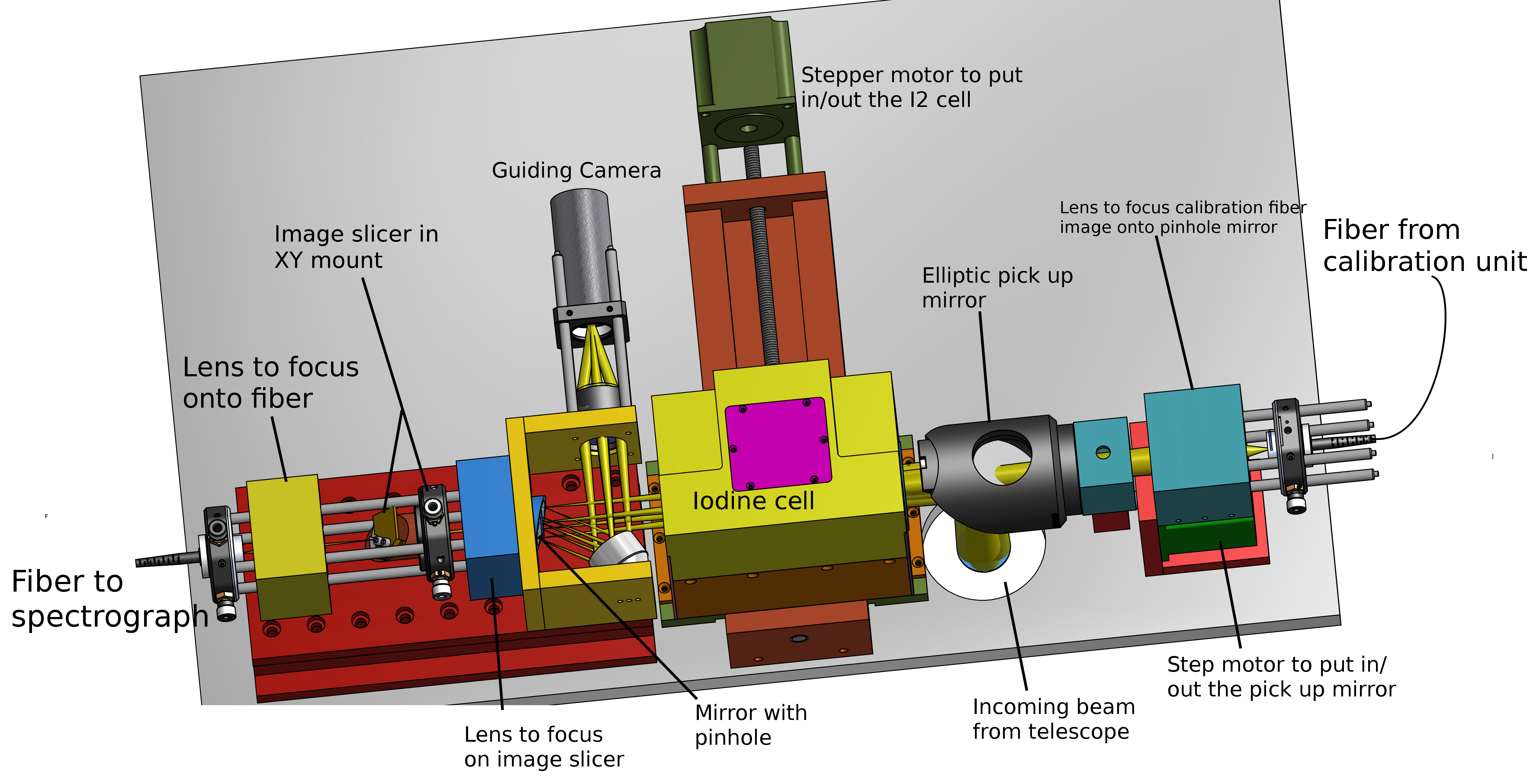}
   \end{tabular}
   \end{center}
   \caption[Front end] 
   { \label{fig:frontend} Opto-mechanical design of the front end module. The size of the plate onto which the opto-mechanical components are mounted is 640$\times$400\,mm. It will be installed on the Nasmyth focus of the Waltz telescope. The beam coming from the telescope is reflected to the iodine cell and then to the mirror-pinhole. The star to be observed goes through the pinhole, is focused by a lens on the image slicer, and then focused onto the spectrograph fiber. When taking calibrations the pick up mirror can me moved out of the telescope beam, letting pass the light coming from the calibration fiber which is aligned with the optical axis of the front end optics.}
   \end{figure} 
   
\begin{figure} [ht]
   \begin{center}
   \begin{tabular}{c} 
   \includegraphics[height=10cm]{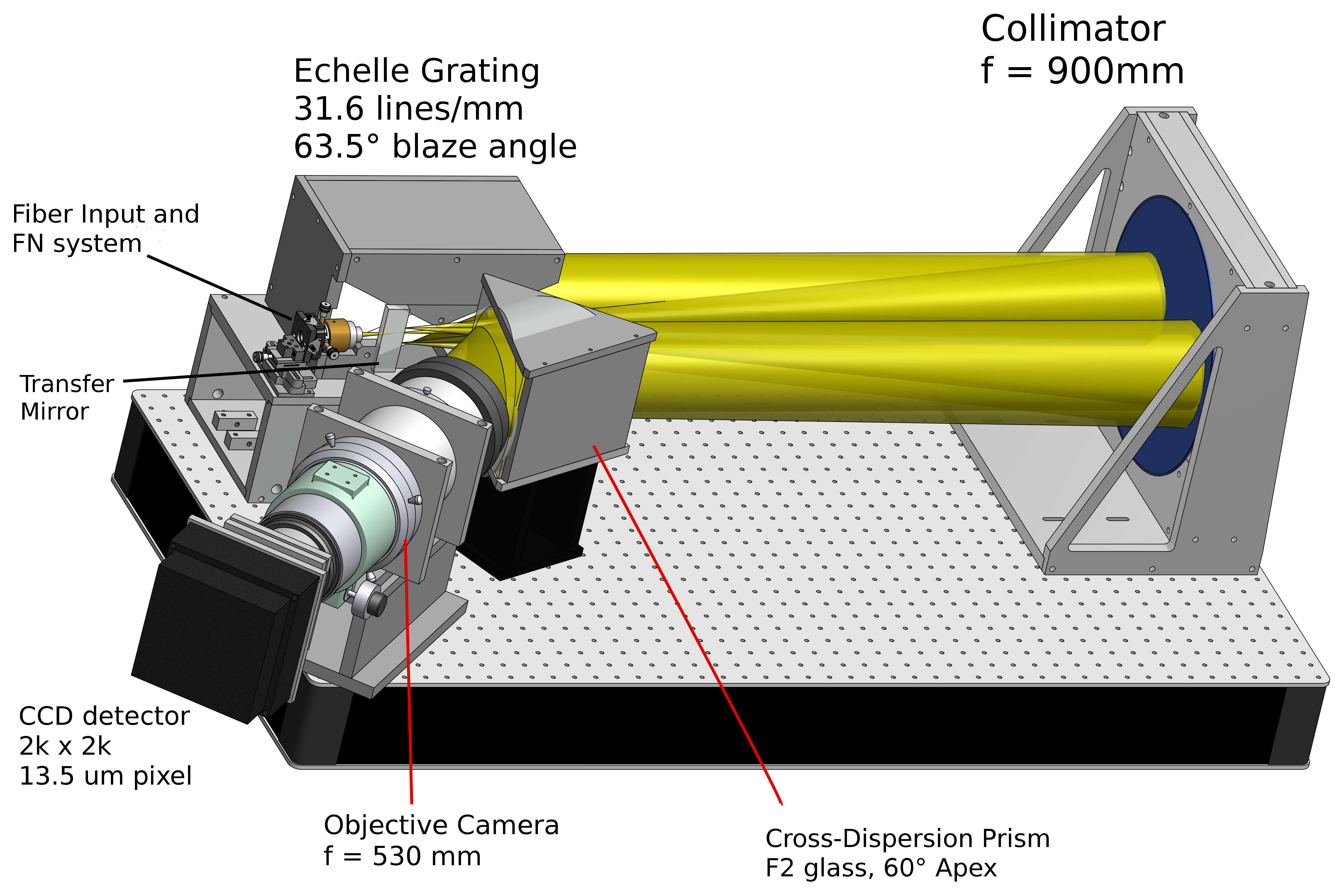}
   \end{tabular}
   \end{center}
   \caption[example] 
   { \label{fig:opto-mech-waltz} Opto-mechanical design of the Waltz Spectrograph.}
   \end{figure} 
\subsection{Front End}
The Waltz Spectrograph is a fiber-fed instrument located in the coude room, and connected by an optical fiber to the telescope focus. The interface for light injection into the fiber is attached to the telescope's Nasmyth focal station. This front end will allow to: 
focusing the light of the observed object into the optical fiber with a suitable focal ratio, injecting light from the calibration unit into the spectrograph, observing the focal plane of the telescope with a guiding camera and moving the iodine cell in and out of the beam for precise RV measurements.\\
We place a mirror with a 200$\upmu$m pinhole tilted by 15$^\circ$ with respect to the telescope optical axis in the focal plane of the telescope. This mirror redirects the image of the field to the guiding camera, a 752$\times$580 pixels CCD detector with 8.2$\times$8.4\,$\upmu$m pixel size, which observes a 2$\times$2 arcmin field of view. The star to be observed goes through the pinhole to an image slicer that transforms the slightly elliptical image of the tilted pinhole into a pseudo-slit image that is injected into the rectangular fiber that feeds the spectrograph{\cite{Spronck2012}}. We adopt a Bowen-Walraven design based on mirrors{\cite{Schwab2010}} for the image slicer. The rectangular fiber FC connector is mounted on a XYZ-translation mount for precise alignment.\\Calibration spectra can be taken by moving out the pick up mirror, letting the light from the calibration fiber, which is aligned with the optical axis of the front end optics, to go through the mirror pinhole, after which is then focused onto the spectrograph fiber. We use a 100$\upmu$m diameter optical fiber with an achromatic lens to transform the calibration fiber output beam into the telescope focal ratio and focus the image of the calibration fiber into the mirror's pinhole. The opto-mechanical design is show in Fig.\ref{fig:frontend}.\\
The front end module also comprises the iodine cell{\cite{Seemann2016}}, built by the University of G\"ottingen, which is located on a remotely movable stage. The cell intercepts the f/18 telescope beam immediately after the Nasmyth folding mirror, and before the focal plane.
The iodine cell is designed for excellent thermal stability and minimal heat dissipation, dictated by its vicinity to the telescope's primary mirror unit. An encapsulated housing was developed that keeps the outside cell structure roughly at ambient temperature even when the cell is heated for long periods of time. 
A rigid mechanical mount provides both mechanical and thermal stability as well as protection upon exposure to the open dome environment during normal operations. A closed loop control system minimizes the overhead during the warm-up phase until stabilization, and monitors the cell's temperature. At $\sim$50$^{\circ}$C, the system provides deep absorption bands of iodine in the entire 500$-$600\,nm range. The iodine spectrum is imprinted onto the stellar spectrum during the observations, and shares the very same optical path as the stellar spectrum along the instrument's optical train. Comparison of the iodine spectral lines with a high-resolution template spectrum of the cell yields precise information on the line spread profile, tracking instrumental instabilities. The observed stellar spectrum with iodine lines imprinted will be fitted with a convolution of a stellar template taken without the iodine cell, and a template with extremely high resolution of the iodine cell itself, yielding the precise velocity shift between the stellar template and the stellar observation with the iodine cell. We use a high-SNR template spectrum of the cell with $R\sim 10^{6}$, obtained with the G\"ottingen Fourier-Transform Spectrometer. 

\subsection{Calibration Unit}
\label{sec:calunit}
The calibration unit contains a Thorium-Argon (Th-Ar) lamp for wavelength calibration and focusing, and a quartz Tungsten-Halogen lamp with high efficiency in the optical for order tracing and flat fielding. Th-Ar lamps provide a dense forest of spectral emission lines in the optical wavelength range with strong Ar emission lines above 650\,nm. The optical axis of the continuum lamp assembly is perpendicular to the Th-Ar lamp and output fiber optical axis. A beam-splitter redirects the continuum light to the achromatic lens focusing at the output fiber of the calibration unit. Achromatic lens and fiber FC connector are mounted on a cage system for ease of alignment.

\subsection{The Spectrograph}
The Waltz Spectrograph is based on an \'echelle design with prism cross-dispersion and white-pupil configuration, working in quasi-Littrow mode. The spectrograph is fed by a 25$\times$100\,$\upmu$m rectangular fiber. It covers the wavelength range 450 $-$ 800\,nm with a resolving power of 65,000 and a mean sampling of 2.6 pixels. The optical layout is shown in Fig.\,\ref{fig:opto-mech-waltz}.\\
The output of the rectangular fiber is imaged into the focus of the collimator. The beam emerges from the core at F/4.6 and it is transformed by the FN-system into the working F/11 of the spectrograph. As a result the fiber image is magnified to a slit with size 52\,$\upmu$m$\times$224\,$\upmu$m. We use an off-axis sub-aperture of a larger parabolic mirror, from SpaceWalk Instruments, to collimate the beam. The mirror is made on a supremax substrate and has 300\,mm diameter and 900\,mm focal length. The collimated beam size is 90\,mm. We choose a relatively small off-axis angle of 4.8 degrees, to keep the focal ratio of the main mirror moderate, and hence easy to manufacture. 
The main dispersion element is an off-the-shelf \'echelle grating with 31.6 lines/mm and blaze angle of 63.5$^{\circ}$ from Richardson Grating Labs. The size of the grating is 200mm$\times$100\,mm$\times$35\,mm, being completely under-filled with the collimated beam. Due to the white-pupil design of the spectrograph, the beam is reflected back to the collimator after the \'echelle dispersion. After the second reflection on the collimator the beam is focused on the transfer mirror, a flat aluminum-coated mirror that reflects the beam back to the parabolic mirror, for a second collimation. The cross-dispersion is provided by a 130\,mm tall prism made of Schott F2 glass with 60$^\circ$ apex angle working in minimum deviation. All prism surfaces have been coated with a silica Sol-Gel anti-reflection coating optimized for the 55$^\circ$ incident angle. Prism characteristics were chosen to provide a minimum order separation of 25 pixels between the reddest orders, to prevent contamination between adjacent orders.\\
For the objective camera we use a modified Petzval quadruplet apochromat with an aperture of 106\,mm and a focal length of 530\,mm (F/5), from Takahashi Corporations (FSQ106 ED). It is optimized for wide-field imaging, providing a high quality image over the entire spectrum.\\
As a detector we will use Andor's iKon-L 936 back-illuminated CCD detector, a 2048$\times$2048 pixel array with 13.5\,$\upmu$m pixel size. Read-out noise and dark current values are 2.9\,\mbox{e$^{-}$} and 0.0004\,\mbox{e$^{-}$/pixel/sec} at -70$^{\circ}C$, respectively.\\
As an exposure meter we will use a photon counter from Hamamatsu Corporations model H9319-11, which will observe the zeroth order of the echelle spectrum. The photon count sensitivity at $\lambda$\,=\,500\,nm is 3.4$\times$105\,s$^{-1}$\,pW$^{-1}$.
  
 \begin{figure} [ht]
   \begin{center}
   \begin{tabular}{c} 
   \includegraphics[height=12cm]{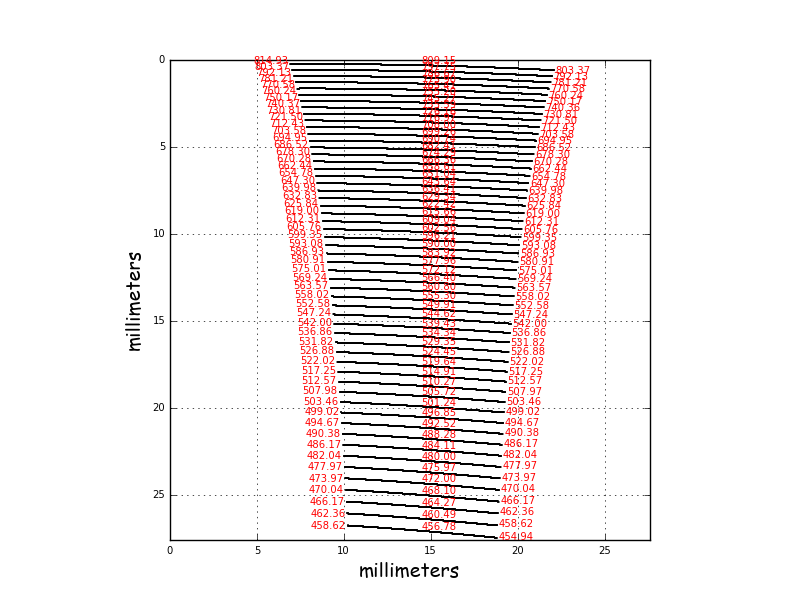}
   \end{tabular}
   \end{center}
   \caption[echellogram] 
   { \label{fig:echellogram} Waltz Spectrograph echellogram. The box is the size of the CCD. Only the free spectral range is shown for each order. The central and extreme wavelengths in nm of each order are shown in red.}
   \end{figure} 

The optical axis height is 192\,mm above the optical bench. The opto-mechanical layout of the spectrograph is shown in Fig.\ref{fig:opto-mech-waltz}, while the Waltz Spectrograph echellogram is presented in Fig.\ref{fig:echellogram}. Its main characteristics are summarized in Table\,\ref{tab:parameters}. All reflective elements are mounted on mechanical mounts that provide small tilt adjustments for precise alignment of the optics. There are no movable parts in the spectrograph unit other than the CCD shutter. The optical bench is a 1200$\times$600\,mm rectangular steel breadboard with M6 mounting holes and excellent thermal stability. All the elements of the spectrograph are mounted directly onto the optical bench except for the detector, which is connected to the objective camera through a special mount that allows precise alignment, as shown in Fig.\,\ref{fig:opto-mech-waltz}. The bench is mounted on legs with a damping system using compressed air to compensate for small mechanical vibrations from the environment, optimizing instrument stability. The spectrograph will be housed in a metallic enclosure that is further isolated by foam panels. Even if there is no temperature control of the instrument, the enclosure will reduce temperature variations. We expect variations on the order of $\pm$5$^{\circ}C$ in the spectrograph unit within a period of two months, based on temperature measurements in the spectrograph room.

\section{Early results and Instrument performance}
\begin{figure} [ht]
   \begin{center}
   \begin{tabular}{c} 
   \includegraphics[height=5cm, angle = 90]{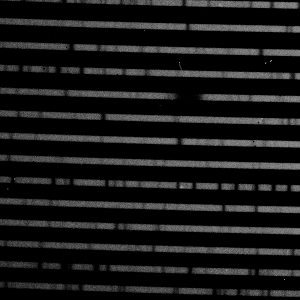}\includegraphics[height=5cm, angle = 90]{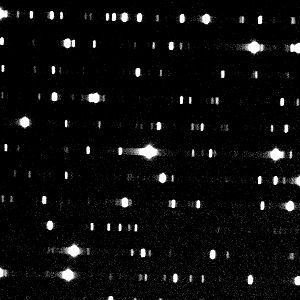}\includegraphics[height=5.5cm]{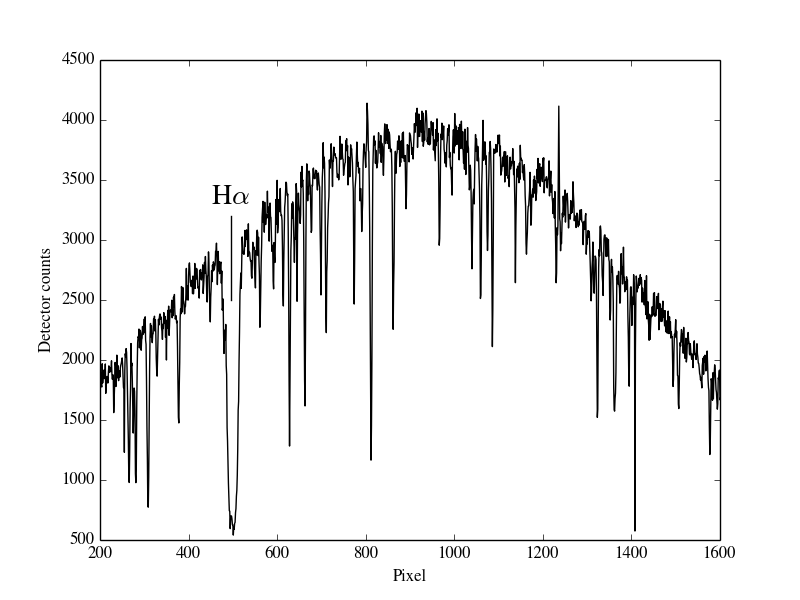}
   \end{tabular}
   \end{center}
   \caption[] 
   { Left: Raw Arcturus and ThAr emission spectrum. Right: Extracted Arcturus spectrum showing H$\alpha$.\label{fig:frames}}
   \end{figure}

We obtained a set of continuum and Th-Ar lamp spectra in our optics laboratory to characterize the spectrograph. We used a 4096$\times$4096 pixel detector with 9$\upmu$m pixel size detector for the lab tests, as the spectrograph CCD is not yet available. We used the reduction and analysis facility IRAF for data reduction. The reduction procedure includes calibration of CCD data (bias, dark and flat-field), extraction of spectra and wavelength calibration. Fig. \ref{fig:frames} shows raw Arcturus and Th-Ar frames obtained with the Waltz Spectrograph.\\
The theoretical resolving power of the Waltz spectrograph is $\sim$65,000 in the optical wavelength range. We used Th-Ar lines both to determine the instrumental dispersion and to measure the resolving power ($R$ = $\lambda$/$\Delta\lambda$) of the spectrograph. A set of Th-Ar lines in the range between 609.88\,nm and 610.56\,nm was used for this purpose. The dispersion was determined by measuring the FWHM of each line in \AA{} of the wavelength calibrated spectra, and dividing it by the FHWM of each line in pixels on the reduced spectra. We obtained an average value of 0.025 $\pm$0.002\AA{}/pix for the spectral dispersion in order 97. For the same lines we measure the resolving power by dividing the central wavelength of each line by its FWHM. For this order we obtained an average value of 61656$\pm$1962, which is in agreement with expectations. The scatter on the resolving power is related to the optical properties of the spectrum itself. There is a small defocus factor at the edges of the \'echelle orders, due to the non-compensated field curvature of the white pupil relay. Spectral lines close to the center of the detector, where the focus is optimal, demonstrate a resolving power up to 66\,000.\\In addition, we measure an excellent internal transmission of 48.9\% at 633\,nm at blaze peak, from the fiber to the back of the objective camera. Considering the theoretical quantum efficiency of the Andor CCD detector, which is $\sim$95\% at 633\,nm, the overall spectrograph efficiency will be 46.5\%.
We performed first test observations of Arcturus and very few other objects so far. A raw and a reduced spectrum are shown in Fig.\,\ref{fig:frames}.

\section{Summary}
The Waltz Spectrograph fabrication and parts assembly for the optical and mechanical components is finished, except for the front end, which is currently under construction. Commissioning of the instrument with the final configuration at the 72\,cm Waltz Telescope is scheduled to start in fall 2016. The opto-mechanical design and main characteristics were presented. The instrument is already installed at the telescope and early observations have been made to test its performance. The resolving power is in agreement with expectations.  In conjunction with an iodine cell, the Waltz Spectrograph is designed to achieve RV measurements with a precision of $\sim$5\,m/s.

\acknowledgments 
We acknowledge funding provided by the tuition fee committee of Heidelberg University to improve local teaching infrastructure. This work is supported by a grant from the Ministry of Science, Research and the Arts of Baden-W\"urttemberg (AZ 6865.1) to Sabine Reffert, as well as by the Innovation Fund FRONTIER, part of the Excellence Initiative II at Heidelberg University (ZUK 49/2 5.2.142).
 
\bibliography{report} 
\bibliographystyle{spiebib} 

\end{document}